\documentstyle[12pt]{article}

\makeatletter
\@addtoreset{equation}{section}

\makeatother

\def\beq{\begin{equation}}
\def\eeq{\end{equation}}
\def\req#1{(\ref{#1})}
\def\bea{\begin{eqnarray}}
\def\eea{\end{eqnarray}}
\def\ben{\begin{enumerate}}
\def\een{\end{enumerate}}

\def\ni{\noindent}
\def\nn{\nonumber}
\def\ms{\medskip}
\def\wt{\widetilde}

\def\brr{\begin{array}}
\def\err{\end{array}}
\def\ds{\displaystyle}
\def\ie{i.e.\ }
\def\eg{e.g.\ }

\begin{document}

\hfill UB-ECM-PF 95/7

\hfill March 1995

%\vspace*{1mm}

\begin{center}

{\LARGE \bf
GUTs in curved spacetime: running gravitational constants,
Newtonian potential
and the quantum corrected gravitational equations}

\vspace{1mm}

\hspace*{-8mm}
{\bf Emilio Elizalde$^{a,b,}$}\footnote{E-mail: eli@zeta.ecm.ub.es},
{\bf Carlos O. Lousto$^{c,}$}\footnote{E-mail: lousto@ifae.es},
{\bf Sergei D. Odintsov$^{b,}$}\footnote{On leave of absence from
Tomsk Pedagogical Institute, 634041 Tomsk, Russia.
E-mail: odintsov@ecm.ub.es},
and {\bf August Romeo$^{a,}$}\footnote{E-mail: august@ceab.es}
\vspace{2mm}

$^a$Center for Advanced Studies CEAB, CSIC,
Cam\'{\i} de Santa B\`arbara, 17300 Blanes  \\
$^b$Department ECM and IFAE,
Faculty of Physics, University of  Barcelona, \\
Diagonal 647, 08028 Barcelona, Spain \\
$^c$IFAE, Group of Theoretical Physics, Autonomous University of  Barcelona
08193 Bellaterra, Spain \\

\vspace{6mm}

{\bf Abstract}
\end{center}

The running coupling constants (in particular, the gravitational one)
are studied in asymptotically free GUTs and in finite GUTs in curved
spacetime, with explicit examples. The running gravitational coupling
is used to calculate the leading quantum GUT corrections to the
Newtonian potential, which turn out to be of logarithmic form in
asymptotically free GUTs. A comparison with the effective theory for
the conformal factor ---where leading quantum corrections to the
Newtonian potential are again logarithmic--- is made.
A totally asymptotically free $O(N)$ GUT with quantum higher derivative
gravity is then constructed, using the technique of introducing
renormalization group (RG) potentials in the space of couplings.
RG equations for the cosmological and gravitational couplings in this
theory are derived, and solved numerically, showing the influence
of higher-derivative quantum gravity on the Newtonian potential.
The RG-improved effective gravitational Lagrangian for asymptotically
free massive GUTs is calculated in the strong (almost constant)
curvature
regime, and the non-singular De Sitter solution to the quantum
corrected
gravitational equations is subsequently discussed. Finally, possible
extensions of the results here obtained are briefly outlined.

\vfill

\newpage

\section{Introduction}

The study of the quantum properties of grand unified theories
(GUTs) in
the presence of a strong curvature is quite an interesting issue,
owing
to the different applications that it can have in a number of
situations. The results of this study are important for a
detailed
knowledge of the early universe, in particular for an accurate
discussion of the known models of inflationary universe (see
\cite{28}
for a review) and as a guide in the construction of new models of
this
kind. Moreover, such a study is relevant for a better
understanding of
quantum effects in the vicinity of a black hole ---in particular,
for
instance, for the calculation of quantum corrections to the
black-hole
entropy (a recent discussion can be found in \cite{30}).
Furthermore,
such considerations are of fundamental importance for the
estimation of
the back reaction effect a quantum field has on the geometry of
spacetime (for an earlier discussion, in the free matter case,
see
\cite{29,31}).

The renormalization group (RG, see \cite{WiKo} for an introduction)
turns
out to be very useful in the discussion of GUTs in curved space
(see,
for example, \cite{1}). The first investigations on these topics,
which included the construction of the RG for GUTs in curved
spacetime
(see \cite{7,8}, and \cite{1} for a complete list of references),
have been followed by a lot of activity, where the subject is
considered
under quite different points of view. Among the different
interesting
phenomena which are specific of this theory we can count
curvature-induced asymptotic freedom \cite{26}, asymptotic
conformal
invariance both in asymptotically-free \cite{7,1} and in finite
GUTs
\cite{14}, applications of phase transitions of Coleman-Weinberg
type in inflationary universes \cite{28}, curvature-induced phase
transitions \cite{33}, and so on.

In the present paper we study the renormalization group properties
of GUTs in curved spacetime. We start from asymptotically free
GUTs in curved spacetime and write the whole system of RG equations
for all the gravitational couplings. Their behaviour is not
asymptotically free, of course. Concentrating mainly on the gravitational
coupling constant $G$, we give its general running form in asymptotically
free GUTs and provide some explicit examples for the gauge groups
$SU(2)$, $E_6$ (Sect. 2). We also discuss the running couplings for
finite GUTs in curved spacetime. The running gravitational coupling is
calculated explicitly and it is shown that quantum corrections to $G$
have an exponential form, unlike in asymptotically free GUTs, where
they behave power-wise (Sect. 3).

Sect. 4 is devoted to the use of  the running gravitational constant
for calculations of radiative corrections to the Newton potential.
In Sect. 5, in order to study how QG effects may change the qualitative
picture obtained in the previous section, we discuss the effective theory
of conformal gravity by Antoniadis and Mottola. This theory aims at
the description of IR quantum gravity. The running gravitational constant
in such a theory looks qualitatively similar to that in asymptotically free
GUTs, but different from the one in the Einstein theory, where quantum
corrections to the Newtonian potential have also been calculated
recently \cite{17}.

Sect. 6  is devoted to the construction of a totally asymptotically free
theory of matter with $R^2$-gravity. Considering an $O(N)$ gauge theory  with
one multiplet of scalars and two multiplets of spinors as matter, and
making use of
the very interesting technique of introducing potentials in RG-coupling space
(similar to a $c$-function), we explicitly construct the regimes where the
total
matter-QG theory is asymptotically free (such study is carried out
numerically).
Then, in Sect. 7 we compare the behaviour of the running gravitational
coupling in this matter-QG system with the running of $G$ in the same theory
with QG being classical. Sect. 8 is devoted to the study of quantum
corrected gravitational equations in the strong curvature regime. The
non-singular De Sitter solution of these equations with GUT corrections
is discussed, as well as a solution of wormhole type. In the
concluding section we summarize our results and outline some possible
extension of the same.

\section{Asymptotically-free GUTs in curved spacetime and the
gravitational coupling constant}

Our considerations start from a
specific GUT in curved spacetime, given by the following
Lagrangian (a
multiplicatively renormalizable one, \cite{1})
\bea
L &=& L_m + L_{ext}, \nn \\
L_m &=& L_{YM} + \frac{1}{2}
(\nabla_\mu \varphi )^2 +
\frac{1}{2} \xi R \varphi^2
 - \frac{1}{4!} f \varphi^4
 - \frac{1}{2} m^2 \varphi^2
+i\overline{\psi} (\gamma^\mu \nabla_\mu-h \varphi ) \psi, \nn \\
L_{ext} &=& a_1R^2 + a_2 C^2_{\mu\nu\alpha\beta} +a_3 G + a_4
\Box R +
\Lambda - \frac{1}{16\pi G}R.
\label{531}
\eea
With an appropriate gauge group, the theory defined by the
Lagrangian
(\ref{531}) contains  gauge fields $A_\mu$, scalars $\varphi$ and
spinors $\psi$, in some representation of the given gauge group.
As usual, the Lagrangian of the external fields must be added
to $L_m$
in order to obtain a theory which is multiplicatively
renormalizable in
curved spacetime \cite{1}. In (\ref{531}) the cosmological constant has
been chosen in a specific form which will be  convenient in order to
facilitate the discussion of the cosmological applications below.

The detailed consideration of the renormalization structure and
the
RG equations for an asymptotically free GUT of the form
(\ref{531}),
based on the gauge groups SU(5), SU(2), O(N), E$_6$ (and some others)
can be found in
the book \cite{1}, where relevant references are listed too. Now,
since
the theory is multiplicatively renormalizable, the effective
Lagrangian
that corresponds to the classical theory (\ref{531}) fulfills the
standard RG. We shall assume that the background fields vary
slowly with respect to the effective mass of the theory and, therefore,
the derivative
expansion technique can be used in order to obtain the effective
action
of the theory. Then, we can restrict our considerations to the
terms
with only two derivatives with respect to the scalar fields and
four derivatives with respect to the purely gravitational terms,
and it
turns out that the structure of the effective Lagrangian just
mimics
the structure of the classical Lagrangian (\ref{531}).

The RG equations satisfied by the effective Lagrangian are:
\beq
{\cal D} L_{eff} = \left( \mu \frac{\partial}{\partial \mu} +
\beta_i \frac{\partial}{\partial \lambda_i} -
\gamma_i \phi_i \frac{\partial}{\partial \phi_i} \right) L_{eff}
(\mu, \lambda_i, \phi_i )=0,
\label{532}
\eeq
where $\mu$ is the mass parameter, $\lambda_i
=(g^2,h^2,f,m^2,\xi,
\Lambda,G,a_1,a_2,a_3,a_4)$ is the set of all coupling constants,
the
$\beta_i$ are the corresponding beta functions, and $\phi_i
=(A_\mu,
\phi, \psi )$ are the fields.
Note that the dependence on the external gravitational field
$g_{\mu \nu}$ is not explicitly shown in $L_{\mbox{eff}}$.

The solution of Eq. (\ref{532}) by the method of the
characteristics
gives (for a similar discussion in the case of the RG improved
Lagrangian in curved space, see \cite{27}):
\beq
L_{eff} (\mu, \lambda_i, \phi_i ) =
L_{eff} (\mu e^t , \lambda_i (t) , \phi_i (t)),
\label{Leff533}
\eeq
where
\bea
\frac{d\lambda_i (t)}{dt} &=& \beta_i (\lambda_i (t)), \qquad
\lambda_i (0)= \lambda_i, \nn \\
\frac{d\phi_i (t)}{dt} &=& -\gamma_i (t)\phi_i (t), \qquad
\phi_i (0)= \phi_i.
\label{534}
\eea
The physical meaning of (\req{Leff533}) and (\req{534}) is that the
effective Lagrangian $L_{eff}$ (called sometimes the Wilsonian
effective action \cite{WiKo}) is found provided its functional form
at some value of $t$ is known (usually the classical Lagrangian
serves as boundary condition at $t=0$). We will come back to the
discussion of $L_{eff}$ later on and, for the moment, we shall
concentrate on the scaling dependence of the coupling constants,
Eq. (\ref{534}).
Note that only using the RG-improvement procedure can one also get
the non-local effective action, which was discussed in Refs.
\cite{PaTo,CaHu} by direct one-loop calculation.

We consider a typical asymptotically free GUT in curved
spacetime. For studying the scaling dependence of the coupling
constants, the RG parameter will be chosen to be $t=\ln (\mu
/\mu_0)$, as usually, where $\mu$ and $\mu_0$ are two different
mass scales (see \cite{1} for a rigorous discussion of the RG in
curved spacetime). The running coupling constants corresponding
to asymptotically free interaction couplings of the theory
have the form
\bea
&& g^2(t) =g^2 \left( 1 + \frac{B^2g^2t}{(4\pi)^2} \right)^{-1},
\ \ \ \ g^2(0)=g^2, \nn \\
&& h^2(t) =\kappa_1g^2(t) , \ \ \ \ \ \
f(t) =\kappa_2g^2(t),
\label{535}
\eea
where $\kappa_1$ and  $\kappa_2$ are numerical constants defined
by the specific features of the theory under consideration (see
\cite{3}-\cite{5} for explicit examples of such GUTs in flat
space, and \cite{1} for a review). As one can see, asymptotic
freedom ($g^2(t) \rightarrow 0$, $t \rightarrow \infty$) is
realized \cite{6}, for the gauge coupling and for the running
Yukawa and scalar couplings.

The study of this kind of asymptotically free GUT in curved
spacetime was been started in Refs. \cite{7,8} (for a review see
\cite{1}), where the formulation in curved space was also
developed. Using the results in those works it is easy to show
that in such theories (for
simplicity, with only one massive scalar multiplet), one obtains
\beq
\xi (t) =\frac{1}{6} + \left( \xi -  \frac{1}{6} \right) \left( 1 +
\frac{B^2g^2t}{(4\pi)^2} \right)^b, \ \ \ \
m^2(t) = m^2\left( 1 + \frac{B^2g^2t}{(4\pi)^2} \right)^b,
\label{536}
\eeq
being $\xi (0) =\xi$, $m^2(0)=m^2$, and where for the different
models the constant $b$ can be either positive \cite{7,8,1},
negative \cite{7,1}, or zero \cite{1}. Notice that the constant
one-loop running coupling $\xi (t)=\xi$ (i.e. $b=0$) usually
corresponds to supersymmetric asymptotically free GUTs, and also that
classical scaling dimensions are not included in the RG equations for
couplings with mass dimension, as usually happens in the RG
improvement procedure.

Turning now to the gravitational coupling constants \cite{1} (see
also \cite{7,8})
\bea
\frac{da_1 (t)}{dt} &=& \frac{1}{(4\pi)^2} \left[ \xi (t) -
\frac{1}{6} \right]^2 \frac{N_s}{2}, \nn \\
\frac{da_2 (t)}{dt} &=& \frac{1}{120(4\pi)^2} \left( N_s+6N_f+
12N_A \right), \nn \\
\frac{da_3 (t)}{dt} &=& - \frac{1}{360(4\pi)^2} \left( N_s+11N_f+
62N_A \right), \nn \\
\frac{d\Lambda (t)}{dt} &=&  \frac{m^4(t) N_s}{2(4\pi)^2} , \nn
\\
\frac{d}{dt} \frac{1}{16 \pi G(t)} &=&  -\frac{m^2(t) N_s}{(4\pi)^2}
\left[ \xi (t) -  \frac{1}{6} \right],
\label{betasaslg}
\eea
where $N_s,N_f$ and $N_A$ are, respectively, the number of real
scalars, Dirac spinors and vectors and we work in euclidean region.
In what follows we shall
consider dynamical spacetimes with constant curvature (as the De
Sitter space), so that the term in $\Box R$ will not appear. As
one can see, the behavior of the gravitational couplings $a_2(t)$
and  $a_3(t)$ is given by $a_{2,3} (t) \simeq a_{2,3}
+\wt{a}_{2,3} \, t$, and it is exactly the same as in the free
matter theory with the same field content. It is not influenced
by the interaction effects.

The most interesting quantity for us will be the gravitational running
coupling constant $G$. As one can easily find from (\req{betasaslg}),
it has the form
\bea
G(t) &=& G_0 \left\{ 1- \frac{16\pi N_s G_0 m^2 (\xi -1/6)}{B^2
g^2
(2b+1)}  \left[ \left( 1 + \frac{B^2g^2t}{(4\pi)^2}
\right)^{2b+1}
-1 \right]\right\}^{-1} \nn \\
 &\simeq & G_0 \left\{ 1+ \frac{16\pi N_s G_0 m^2 (\xi -1/6)}{B^2
g^2 (2b+1)}  \left[ \left( 1 + \frac{B^2g^2t}{(4\pi)^2}
\right)^{2b+1} -1 \right]\right\} \nn \\
 &\simeq & G_0 \left[ 1+ \frac{ N_s}{\pi} G_0 m^2 (\xi -1/6) \, t
\right], \label{538}
\eea
where $t=\ln (\mu /\mu_0)$. As one can see, the matter quantum
corrections to the gravitational coupling constant in GUTs are
larger than in the free matter theory if $b>0$.

Let us now give some examples. First of all, we consider the
asymptotically free SU(2) model of Ref. \cite{3} with a scalar
triplet and two spinor triplets. Then, one can show \cite{7} that
$N_s=3$, $B^2 =10/3$, $b \simeq 2$. In this model (with  the
standard choice $g^2 \simeq 0.41$) we get for the first non-trivial
correction to the classical $G$:
\beq
G(t) \simeq G_0 \left[ 1 + 0.9549 \, G_0 m^2 (\xi
-1/6)\, t \right]. \label{539}
\eeq
As a second interesting example we will consider the
asymptotically
free E$_6$ GUT \cite{9} in curved spacetime \cite{10}. This
theory
contains a 78-plet of real scalars $\phi$ and two 27-plets of
charged scalar, $M$ and $N$. Assuming that only one mass in the
real scalar multiplet is different from zero, and choosing the
initial values for the charged scalars to be $\xi_M =\xi_N = 1/6$
we obtain again a $G(t)$ of the same form as (\ref{538}). With
the
following parameters \cite{10}: $N_s = 78$, $B^2 =32$,
$g^2=0.41$, we get
\bea
\xi_\phi (t) &\simeq& \frac{1}{6} + 0.97  (\xi_\phi -1/6) \left(
1
+  \frac{B^2g^2t}{(4\pi)^2} \right)^{1/2}, \nn \\
m^2_\phi (t) &\simeq& 0.97 m^2_\phi  \left( 1 +
\frac{B^2g^2t}{(4\pi)^2} \right)^{1/2}.
\label{5311}
\eea
Then, we obtain
\bea
G(t) &=& G_0 \left\{ 1+ \frac{16\pi N_s G_0 m_\phi^2 0.97^2
(\xi_\phi -1/6)}{B^2 g^2 (2b+1)}  \left[ \left( 1+
\frac{B^2g^2t}{(4\pi)^2}
\right)^{2b+1}-1\right] \right\} \nn \\
 &\simeq & G_0 \left[ 1+ \frac{N_s}{3} \, 0.97^2 \, G_0 m_\phi^2
(\xi_\phi -1/6) \, t \right] \nn \\
 &\simeq & G_0 \left[ 1+ 0.8985 \, G_0 m_\phi^2 (\xi_\phi
-1/6)\, t \right]. \label{5312}
\eea
We thus see that we are able to obtain the general form  of the
scaling gravitational coupling constant in an asymptotically free
GUT in curved spacetime. As we observe from these explicit
examples, there exist asymptotically free GUTs in which the
matter
quantum corrections to the gravitational coupling constants at
high
energies are bigger than these corrections in the free matter
theory. The
application of the above results will be discussed later. The
gravitational constants for other asymptotically free GUTs can be
obtained in a similar way.
\ms

\section{Finite GUTs in curved spacetime and the gravitational
coupling constant}

A very interesting class of GUTs in curved spacetime is given by
finite GUTs (see, for instance, \cite{11}-\cite{13}). Finite GUTs
in curved spacetime were first considered in Ref. \cite{14} (for
a
review and a complete list of references, see again \cite{1}).
Of course, although finite for the interaction couplings in flat
space, these theories are not finite in curved spacetime, due to
vacuum, mass and $R\varphi^2$-type divergences.
 Recently, quite a spectacular development has emerged which
concerns $N=2$ supersymmetric models (in particular, finite) with
matter multiplets, through the definition of the exact spectrum
\cite{23}.  For such theories we have (in the following we
consider
one-loop finite supersymmetric or non-supersymmetric theories)
\beq
g^2(t)=g^2, \ \ \ \ h^2(t)=\kappa_1 g^2, \ \ \ \ f(t) =\kappa_2
g^2(t), \label{5313}
\eeq
and
\beq
\xi (t) = \frac{1}{6} + \left( \xi - \frac{1}{6} \right) \exp
(Cg^2t), \ \ \ \ \ \ m^2(t) = m^2  \exp (Cg^2t),
 \label{5314}
\eeq
where the constants $\kappa_1$ and $\kappa_2$ depend on the
specific features of the theory, and where $C$ can be positive,
negative or zero \cite{14}. Solving the RG equations for the
gravitational coupling constant in such theory, we find
\bea
G(t) &=& G_0 \left[ 1- \frac{N_s G_0 m^2 (\xi -1/6)}{2\pi C g^2 }
\left( e^{2Cg^2t} -1 \right)\right]^{-1} \nn \\
 &\simeq & G_0 \left[ 1+ \frac{ N_s G_0 m^2 (\xi -1/6)}{2\pi C
g^2
}   \left(  e^{2Cg^2t} -1 \right)\right]. \label{5315}
\eea
Hence, as one can easily see, GUTs with a positive $C$ are the
ones
which give the biggest contribution to the gravitational coupling
constant, among all finite GUTs.

As an example, one can consider the SU(2) finite model of the
first
of Refs. \cite{13},  with a SU(4) global invariance and the
scalar
taken in the adjoint representation of SU(2). (Notice that there
is
$N=2$ supersymmetry here, in one of the regimes of finiteness.)
In
this case, we get that \cite{14} $N_s=18$ and $C=24/(4\pi)^2$. As
a result,
\beq
G(t)= G_0 \left[ 1+ 6 \pi \, \frac{G_0 m^2 (\xi -1/6)}{g^2
} \left( e^{3g^2t/\pi^2}-1\right) \right]
\label{5316}
\eeq
As one can see, we have power corrections ---of the form
$\sim \left( \mu \over \mu_0 \right)^a$--- to the gravitational
coupling constant.
In a similar way one can also obtain the running gravitational coupling
corresponding to other finite GUTs.
\ms

\section{Quantum GUTs corrections to the Newtonian potential}

As an application of the results of the previous discussion, we
will consider in this section the quantum corrections to the
gravitational potential. From the discussion in Sects. 2 and 3 we
know that the typical behavior of the gravitational coupling
constant, when taking into account the quantum corrections is
\beq
G(t) \simeq G_0 \left\{ 1 + \wt{G}_0 \left[ (1 + \wt{B} t
)^{2b+1}
-1 \right] \right\},
\label{5317}
\eeq
in asymptotically free GUTs, and
\beq
G(t) \simeq G_0 \left[ 1 + \wt{G}_0 \left( e^{\wt{C} t} -1
\right)
\right],
\label{5318}
\eeq
in finite GUTs. Of course, such corrections are too small to be
measured explicitly, although they are significantly larger than
in
free matter theories (due to running). In the above scaling
relations, $ t =\ln (\mu /\mu_0)$.

It was suggested in Refs. \cite{15,16} that in the running
gravitational and cosmological constants $\mu/\mu_0$ ought to be
replaced with $r_0/r$, i.e. that one should change the mass scalar ratio
by the inverse ratio between distances. The support for this
argument
comes from quantum electrodynamics, where the well-known
electrostatic potential with quantum corrections can be
alternatively obtained from the classical potential by
interchanging   the classical electric charge with the running
one,
with $t  =\ln (r_0 /r)$.
Similarly, one can estimate now the GUTs quantum corrections  to
the gravitational potential. Starting from the classical Newtonian
potential
\beq
V(r) = - \frac{Gm_1m_2}{r},
\label{5319}
\eeq
the classical gravitational constant in (\ref{5319}) is to be
changed with the running gravitational coupling constant
(\ref{5317}), (\ref{5318}).

As a result, we obtain,
\beq
V(r) \simeq - \frac{G_0 m_1m_2}{r} \left[ 1+ c_1 G_0 m^2  \left( \xi
- \frac{1}{6} \right) \, \ln \frac{r_0}{r}
\right],
\label{5320}
\eeq
in asymptotically free GUTs, and
\beq
V(r) \simeq - \frac{G_0 m_1m_2}{r} \left[ 1+ c_2 G_0 m^2  \left( \xi
- \frac{1}{6} \right) \left( r_0\over r \right)^{Cg^2}
\right],
\label{5321}
\eeq
in finite GUTs. Here $G_0$ is the initial value of $G$ (the value
at distance $r_0$), and $C_1$ and $C_2$ are some constants.

In Ref. \cite{17} the gravitational potential has been calculated
in the frame of quantum Einsteinian gravity considered as an
effective theory (this is unavoidable, owing to its well-known
non-renormalizability \cite{18}). It was found that in such
theory the leading quantum corrections are proportional to
$G_0/r^2$. As we see, in an asymptotically free GUT the leading
correction is of logarithmic form $\sim \ln (r_0/r)$
while in a finite GUT it is power-like $\sim (r_0/r)^{Cg^2}$.
Hence, from the examples of Sects. 2 and 3 one understands that
the GUT quantum corrections to the Newtonian potential can be
more important than the corresponding corrections in the
effective theory of Einstein's quantum gravity. We now turn to
the study of the gravitational constant in some model of quantum
gravity.
\ms

\section{The gravitational constant in the effective theory for
the conformal factor}

In order to see how quantum gravity (QG) effects change the
results of the above discussion, we have to consider some
multiplicatively renormalizable QG (with matter). For example, in
multiplicatively renormalizable pure higher derivative gravity (for a
review, see \cite{1}) the running gravitational coupling has been
used to construct the QG corrected Newtonian potential in Ref.
\cite{15}. It has been suggested there that the QG corrections in
the Newtonian potential (which are also of logarithmic form)
might help to solve the dark matter problem and some other cosmological
problems \cite{34}.

In this section we are going to discuss the running gravitational
coupling constant in the effective theory for the conformal
factor \cite{19}, which presumably describes the infrared phase
of QG. The construction of the effective theory for the conformal
factor proceeds as follows \cite{19}. One starts from the
conformally invariant matter theory (the free theory, for
simplicity) in curved spacetime. The standard expression for the
conformal anomaly is known (see \cite{21}, and for reviews
\cite{31,1}), and it can be integrated on a conformally-flat
background (in the conformal parametrization $g_{\mu\nu}=
e^{2\sigma} \bar{g}_{\mu\nu}$), in order to construct the
anomaly-induced action. Adding to this action the classical
Einsteinian action in the conformal parametrization, we get the
effective theory for the conformal factor. In the notations of
Ref. \cite{19} and for the flat background $\bar{g}_{\mu\nu}=
\eta_{\mu\nu}$, the resulting Lagrangian is
\beq
L= - \frac{Q^2}{(4\pi)^2} (\Box \sigma )^2 - \zeta \left[ 2\alpha
(\partial_\mu \sigma )^2 \Box \sigma  + \alpha^2
(\partial_\mu \sigma )^4 \right] + \tilde\gamma e^{2\alpha\sigma}
(\partial_\mu \sigma )^2 -\frac{\lambda}{\alpha^2}
e^{4\alpha\sigma},
\label{5322}
\eeq
where $\tilde\gamma =3/(8 \pi G)$, $\lambda = \Lambda/(8\pi G)$, $\zeta
=b+2b'+3b''$ and $Q^2/(4\pi)^2 =\zeta- 2b'$, being $b,b'$ and
$b''$ the well-known coefficients of the conformal anomaly
\cite{21} (for a recent discussion, see \cite{22}):
\beq
T_\mu^{\ \mu} = b \left(C_{\mu\nu\alpha\beta}^2 + \frac{2}{3}
\Box R \right) + b' G + b''\Box R.
\label{5323}
\eeq
$Q^2$ has been interpreted as the 4-dimensional central charge
and $\alpha^2$ is the anomalous scaling dimension for $\sigma$.

Near the infrared (IR) stable fixed point $\zeta =0$, the theory
(\ref{5322}) has been argued to describe the IR phase of QG
\cite{19}. Near the IR fixed point $\zeta =0$, the inverse of the
running gravitational coupling constant in the IR limit ($t
\rightarrow -\infty$) has been calculated in Ref. \cite{20}:
\beq
\tilde\gamma (t) \simeq (-t)^{-1/5} \exp \left[ t \left(2-2\alpha +
\frac{2\alpha^2}{Q_0^2} \right) \right],
\label{5324}
\eeq
where $Q_0^2=Q^2 (\zeta =0)$. As a result, we obtain the
Newtonian potential with account to the quantum corrections in
such a model, as \cite{20}
\beq
V(r) = - \frac{G_0 m_1m_2}{r} \left( 1- \frac{\alpha^2}{10Q_0^2} \ln
\frac{r_0^2}{r^2} \right).
\label{5325}
\eeq
As we can see here, essential leading-log corrections to the
Newtonian potential appear in this model, what is different from
what happens in the case of Einstein gravity.
\ms

\section{Asymptotic freedom in GUTs interacting with
higher derivative quantum gravity}

After the above discussion on the running gravitational coupling
constant, mainly for GUTs in curved spacetime, our purpose will be to
look at its quantum gravitational corrections in frameworks of
asymptotically free theories. To be specific, we start from a model
with the following Lagrangian (here we use notations slightly different from
those in Sect. 2)
\beq
\begin{array}{lll}
{\cal L}&=&\ds{1 \over 2\lambda}
C_{\mu \nu \alpha \beta} C^{\mu \nu \alpha \beta}
-{\omega \over 3 \lambda} R^2
-{1 \over 4}G^a_{\mu \nu} G^{a \mu \nu} \\
&&\ds+{1 \over 2} g^{\mu \nu}
 ({\cal D}_{\mu}\varphi)^i ({\cal D}_{\nu}\varphi)^i
+{1 \over 2} \xi R \varphi^i \varphi^i
-{1 \over 4!}f ( \varphi^i \varphi^i )^2 \\
&&\ds -{1 \over 2} m^2 \varphi^i \varphi^i
 -{R \over 16\pi G}+ {\Lambda \over 8\pi G} \\
&&\ds + i \bar{\psi}_p
[ \gamma^{\mu}{\cal D}_{\mu}^{pq} -h_i^{pq}\varphi^i ]
\psi_q .
\end{array}
\label{L25}
\eeq
We will not discuss the problem of unitarity in this theory, which
is still open \cite{AnTo} and may be solved, perhaps, only
non-perturbatively.
We rather consider \req{L25}
as an effective theory for some unknown consistent QG.
This model includes in its gravitational sector higher derivative
gravity (with somehow different notations for the gravitational
couplings) and in the matter sector an $O(N)$ gauge theory with scalars
in its fundamental representation plus $n_1$ spinor mutliplets in the
adjoint representation and $n_2$ spinor mutliplets in a fundamental
representation. It is known \cite{5,35} that for some values of
$N, n_1, n_2$ such a flat GUT provides asymptotic freedom for all its
coupling constants. Moreover, even taking into account quantum gravity,
one can show that asymptotic freedom may survive (see \cite{36}) but
under tighter restrictions on the contents of the theory.
When studying the asymptotically free regime we look at the region
$h^2 \ll g^2$ \cite{36} (so that Yukawa coupling corrections may be
dropped). One can show that such a description is consistent under
radiative corrections.
It is interesting to remark that in the low-energy limit the theory
\req{L25} leads to the standard Einstein action \cite{14}
---similarly to what happens in the pregeometry program \cite{Sakh}.

The one-loop beta functions for the theory \req{L25} have been
calculated in \cite{36,1}.
%, and read
%\beq
%\begin{array}{lllll}
%\ds{d\lambda\over dt}&=&\beta_{\lambda}&=&\ds
%-\kappa a^2 \lambda, \\
%\ds{d g^2\over dt}&=&\beta_{g^2}&=&\ds
%-\kappa b^2 g^4, \\
%\ds{d\omega\over dt}&=&\beta_{\omega}&=&\ds
%-\kappa\lambda\left[ {10 \over 3}\omega^2
%+\left( 5+a^2 \right) \omega + {5 \over 12}
%+{3 \over 2}N \left( \xi -{1 \over 6} \right)^2 \right], \\
%\ds{d\xi\over dt}&=&\beta_{\xi}&=&\ds\kappa
%\left[ \left( \xi -{1 \over 6} \right)
%\left( {N+2 \over 3}f-{3\over 2}(N-1)g^2 \right) \right. \\
%&&&&\ds\left.\hspace{1em} +\lambda\xi\left(
%-{3 \over 2}\xi^2 +4\xi +3 +{10 \over 3}\omega
%-{9 \over 4 \omega}\xi^2 +{5 \over 2\omega}\xi- {1 \over 3\omega}
%\right) \right] \\
%&&&\equiv&\beta_{\xi}^{(0)}+\Delta\beta_{\xi}, \\
%\ds{df\over dt}&=&\beta_f&=&\ds
%\kappa \left[
%{N+8 \over 3} f^2-3(N-1)fg^2+{9 \over 4}(N-1)g^4 \right. \\
%&&&&\hspace{1em}\ds\left. +\lambda^2\xi^2\left( 15+ {3 \over 4\omega^2}
%-{9 \xi \over \omega^2}+{27 \xi^2 \over \omega^2} \right) \right. \\
%&&&&\hspace{1em}\ds \left. -\lambda f\left(
%5+3\xi^2+{33 \over 2\omega}\xi^2
%-{6 \over \omega}\xi +{1 \over 2 \omega} \right) \right] \\
%&&&\equiv&\beta_f^{(0)}+\Delta\beta_f,
%\end{array}
%\label{betaslgoxf}
%\eeq
We will be interested in the study of that system of equations in the
$t\to\infty$ (or high-energy) limit, where it is convenient to
make the variable changes \cite{36}
\beq
\begin{array}{lll}
\bar{f}&=&\ds{f \over g^2} , \\
d\tau&=&\ds\kappa g^2(t) dt , \hspace{1cm}
\kappa\equiv{1 \over (4\pi)^2}.
\end{array}
\label{changes}
\eeq
Further, we will take advantage of the property that, in the limit
considered,
$\ds{\lambda(t) \over g^2(t)}\to\ds{b^2 \over a^2}$.
First, we look at the case of massless couplings only. In these
conditions, the total system of RG equations turns into
\beq
\begin{array}{lllll}
\ds{d g^2\over d\tau}&=&\bar\beta_{g^2}&=&\ds-b^2 g^2, \\
\ds{d\omega\over d\tau}&=&\bar\beta_{\omega}&=&\ds
-{b^2\over a^2}\lambda\left[ {10 \over 3}\omega^2
+\left( 5+a^2 \right) \omega + {5 \over 12}
+{3 \over 2}N \left( \xi -{1 \over 6} \right)^2 \right], \\
\ds{d\xi\over d\tau}&=&\bar\beta_{\xi}&=&\ds
\left( \xi -{1 \over 6} \right)
\left( {N+2 \over 3}\bar{f}-{3(N-1)\over 2}g^2 \right) \\
&&&&\ds +{b^2\over a^2} \xi\left(
-{3 \over 2}\xi^2 +4\xi +3 +{10 \over 3}\omega
-{9 \over 4 \omega}\xi^2 +{5 \over 2\omega}\xi- {1 \over 3\omega}
\right) \\
\ds{d\bar{f}\over d\tau}&=&\bar\beta_{\bar{f}}&=&\ds
{N+8 \over 3} \bar{f}^2+(b^2-3(N-1))\bar{f}+{9 \over 4}(N-1) \\
&&&&\ds\left. + \left( b^2\over a^2 \right)^2 \xi^2
\left( 15+ {3 \over 4\omega^2}
-{9 \xi \over \omega^2}+{27 \xi^2 \over \omega^2} \right) \right. \\
&&&&\ds \left. -{b^2\over a^2} \bar{f}\left(
5+3\xi^2+{33 \over 2\omega}\xi^2
-{6 \over \omega}\xi +{1 \over 2 \omega} \right) \right] , \\
\end{array}
\label{betasgoxf}
\eeq
where
\beq
\begin{array}{lll}
\kappa&=&\ds{1 \over (4 \pi)^2} , \\
a^2&=&\ds{1 \over 60}(798+6N^2-5N)
+{N \over 10}\left( n_2+{1 \over 2}(N-1)n_1 \right) , \\
b^2&=&\ds{1 \over 6}(22N-45)-{4 \over 3}( n_1(N-2)+n_2 ) .
\end{array}
\label{defsa2b2}
\eeq
It is evident that asymptotic freedom (AF) in the original
couplings is determined by the existence of stable fixed points
for this new system. Note that the RG equations \req{betasgoxf}
are also very useful for explicit discussions of different forms of
effective potential in quantum matter-$R^2$-gravity theories
\cite{33,EOR}

\subsection{Without gravity}
To make our discussion easier, we start by considering
the same situation when gravity is switched off. In these conditions,
\beq
\begin{array}{lll}
\ds{d g^2\over d\tau}&=&\ds-b^2 g^2, \\
\ds{d\bar{f}\over d\tau}&=&\ds
{N+8 \over 3} \bar{f}^2+(b^2-3(N-1))\bar{f}+{9 \over 4}(N-1).
\end{array}
\label{betasgf}
\eeq
This system will be examined using the methods developed in \cite{37},
where some potentials in RG coupling-space were introduced, so that
their stability properties yield the fixed points of the original system.
One introduces a renormalization group potential $u(g^2, \bar{f})$ such
that
\beq
\begin{array}{lll}
\ds{\partial u \over \partial g^2}&=&\ds{d g^2\over d\tau}, \\
\ds{\partial u \over \partial \bar{f}}&=&\ds{d \bar{f}\over d\tau}.
\end{array}
\label{defugf}
\eeq
With the sign convention here used, the existence of stable fixed points
amounts to the presence of some sort of maximum for $u$ (if the signs
in \req{defugf} were reversed,
we should say minimum instead of maximum).
Up to an arbitrary constant, this potential reads
\beq
u(g^2, \bar{f})=-{b^2\over 2}g^4+
{N+8 \over 9} \bar{f}^3+{(b^2-3(N-1)) \over 2}\bar{f}^2
+{9 \over 4}(N-1)\bar{f}.
\eeq
First, we obtain the values of the fixed points, which correspond to the
critical points of $u$ in ($g^2, \bar{f}$)-space. Trivially,
$g^2=0$, but there are real solutions for $\bar{f}$ only when
\[ \Delta \equiv (b^2-3(N-1))^2-3(N+8)(N-1) \geq 0. \]
If $\Delta \geq 0$, the
critical values of $\bar{f}$ are
\beq
\bar{f}_{1 \atop 2}=
{ -(b^2-3(N-1))\pm\sqrt{\Delta} \over 2/3(N+8) } .
\label{barf12}
\eeq
Therefore, the Hessian matrix at these points is
\beq
\left(
\begin{array}{cc}
\ds{\partial^2 u \over \partial(g^2)^2}&
\ds{\partial^2 u \over \partial g^2\partial\bar{f}} \\
\ds{\partial^2 u \over \partial\bar{f}\partial g^2}&
\ds{\partial^2 u \over \partial\bar{f}^2}
\end{array}
\right)_{ g^2=0, \bar{f}=\bar{f}_{1 \atop 2} }
=
\left(
\begin{array}{cc}
-b^2&0\\
0&\pm\sqrt{\Delta}
\end{array}
\right) .
\eeq
Obviously, its eigenvalues are $b^2$ and $\pm\sqrt{\Delta}$.
We need $b^2 >0$ in order to ensure AF in $g^2$.
Thus, $u$ can have an extreme only when picking the minus sign,
which corresponds to $\bar{f}=\bar{f}_2$, and then that extreme
is a local maximum. In such a set-up AF for $\bar{f}$ takes place.
By the same argument
we conclude that $\bar{f}=\bar{f}_1$ corresponds to a saddle point.
The hypothetical case $\Delta=0$ is
rather exceptional; there is only one solution
for $\bar{f}$ and the outcome is one vanishing eigenvalue
(\ie the Hessian no longer has maximal rank). Actually, while there is a
maximum along the $g^2$ axis, the potential shows just an inflexion
point along the $\bar{f}$ direction, typical of the cubic dependence of
$u$ in this variable.

When considering only the
$\bar{f}$-dependent part of $u$ ---say $u_{\bar{f}}$--- we realize that
it has no global maximum or minimum, because cubic curves are unbounded
from above or below. However,
if $\Delta >0 $, it has a local minimum (at $\bar{f}=\bar{f}_1$) and a
local maximum (at $\bar{f}=\bar{f}_2$). If $\Delta =0 $, it has an
inflexion point at the only zero of $\beta_{\bar{f}}$ and no local
extremes. On the contrary, when $\Delta < 0$, $u_{\bar{f}}$ is a
monotonic curve without extremes or inflexion points. In the end, it is
the presence of the local maximum $\bar{f}=\bar{f}_2$ that accounts for
AF.

As we have just seen, the AF scenario is determined by
$b^2 >0$ and $\Delta > 0$. Whether these constraints can be met or not
depends, of course, on the particular values of $N$ and of $b$ which,
in turn, depends on $N$, $n_1$ and $n_2$.
Assuming $N>2$, and bearing in mind that $n_1, n_2$ are integers
larger than or equal to zero, from \req{defsa2b2} one readily finds
the condition
\beq
n_1 \leq 2 .
\eeq
Then, after numerical examination of $\Delta$ for $1 \leq N \leq 10$,
$0 \leq n_1 \leq 2$ and $0 \leq n_2 \leq 20$, we find the following
combinations yielding AF solution:
\begin{center}
\begin{tabular}{|r|r|l|}
$N$&$n_1$&$n_2$ \\ \hline
 7&0&13 \\
  &1&8  \\
  &2&3  \\
 8&0&15,16 \\
  &1& 9,10 \\
  &2& 3, 4 \\
 9&0&17,18,19 \\
  &1&10,11,12 \\
  &2& 3, 4, 5 \\
10&0&19,20 \\
  &1&11,12,13 \\
  &2& 3, 4, 5 \\ \hline
\end{tabular}
\end{center}
This constitutes an example of (discrete) numerical boundary separating
the region where the theory shows a given behaviour for
$\tau\to\infty$.

Let's now see, from another viewpoint, the meaning of our constraints
$b^2 >0$ and $\Delta > 0$
when looking at the couplings themselves. In the present case, we can
integrate the RG equations \req{betasgf}. For $g^2$ one has
\beq g^2( \tau )= g^2(0) e^{-b^2 \tau}  \label{g2tau} \eeq
(we are taking initial conditions so that $\tau=0$ when $t=0$).
It is plain ---as we have said--- that $b^2>0$ leads to AF in $g^2$,
while $b^2<0$ entails boundless increase of $g$ as $\tau$ grows,
making perturbation theory no longer valid.
With regard to $\bar{f}$, there are three cases depending on $\Delta$:
\begin{enumerate}

\item $\Delta < 0$ (no fixed points)
\beq
\bar{f}(\tau)={3 \over 2(N+8)}
\left[ \sqrt{-\Delta} \tan\left( \sqrt{-\Delta} \tau \over 2 \right)
-(b^2-3(N-1)) \right] .
\eeq
$\bar{f}$ is periodic in $\tau$, having regularly spaced singularities
--- analogous to Landau poles.

\item $\Delta =0$ (one double fixed point)
\beq
\bar{f}(\tau)=-{ (b^2-3(N-1)) \over 2/3(N+8) }
-{1 \over 2/3(N+8)\tau} .
\eeq
%The poles are no longer present but,
%since $|\bar{f}|$ increases linearly without bound, it is obvious that
%this fixed point is unstable.
There is stability in the sense that
$\ds \bar{f}(\tau)\to -{ (b^2-3(N-1)) \over 2/3(N+8) }$ as
$\tau\to\infty$, but we still have a pole of this coupling at $\tau= 0$.

\item $\Delta >0$ (two single fixed points)
\beq
\bar{f}(\tau)={
(b^2-3(N-1)-\sqrt{\Delta}) e^{-\sqrt{\Delta}\tau}
-(b^2-3(N-1)+\sqrt{\Delta}) \over
2/3(N+8) (1-e^{-\sqrt{\Delta}\tau})
}
\label{ftauDeltag0}
\eeq
It is clear that now a finite limit exists when $\tau$ goes to infinity.
In fact,
\beq
\begin{array}{lll}
\bar{f}(\tau)&\to&\ds-{(b^2-3(N-1)+\sqrt{\Delta}) \over 2/3(N+8)}
=\bar{f}_2 \\
\tau&\to&\infty ,
\end{array}
\eeq
as should be expected, because we already knew that $\bar{f}_2$ is the
stable fixed point. Now, there can be no doubt that the previous
criteria are right and this is indeed the AF region.

\end{enumerate}

\subsection{With gravity}

Next, we will have to deal with Eqs. \req{betasgoxf}, which offer more
serious difficulties than the system \req{betasgf}. Since not all the
crossed derivatives coincide, we cannot just integrate and find
a potential which is a function of all the variables. Instead, we may
handle individual potentials for every constant, like in \cite{37}.
If we call them $u_{g^2}, u_{\omega}, u_{\xi}, u_{\bar{f}}$, one has
to require
\beq
\begin{array}{lll}
\ds{\partial u_{g^2} \over \partial g^2}&=&\ds{d g^2\over d\tau}, \\
\ds{\partial u_{\omega} \over \partial\omega}&=&\ds{d \omega\over d\tau},\\
\ds{\partial u_{\xi} \over \partial \xi}&=&\ds{d \xi\over d\tau},\\
\ds{\partial u_{\bar{f}} \over \partial \bar{f}}&=&\ds{d \bar{f}\over d\tau} .
\end{array}
\label{defugoxf}
\eeq
Some possible solutions are
\beq
\begin{array}{lll}
u_{g^2}&=&\ds-{b^2\over 2}g^4, \\
u_{\omega}&=&\ds
-{b^2\over a^2}\lambda\left[ {10 \over 9}\omega^3
+{ 5+a^2 \over 2} \omega^2
+ \left( {5 \over 12}
+{3 \over 2}N \left( \xi -{1 \over 6} \right)^2 \right)\omega \right],
\\
u_{\xi}&=&\ds
\left( {\xi^2 \over 2} -{1 \over 6}\xi \right)
\left( {N+2 \over 3}\bar{f}-{3(N-1)\over 2}g^2 \right) \\
&&\ds +{b^2\over a^2} \left(
-{3 \over 8}\xi^4 +{4\over 3}\xi^3 +{3\over 2}\xi^2
+{10 \over 3}\omega\xi
-{9 \over 16 \omega}\xi^4 +{5 \over 6\omega}\xi^3
-{1 \over 6\omega}\xi^2
\right), \\
u_{\bar{f}}&=&\ds
{N+8 \over 9} \bar{f}^3
+\left[ {b^2-3(N-1) \over 2}
-{b^2\over 2 a^2} \bar{f}\left(
5+3\xi^2+{33 \over 2\omega}\xi^2
-{6 \over \omega}\xi +{1 \over 2 \omega} \right)
\right]
\bar{f}^2 \\
&&\ds + \left[ {9 \over 4}(N-1)
+ \left( b^2\over a^2 \right)^2 \xi^2
\left( 15+ {3 \over 4\omega^2}
-{9 \xi \over \omega^2}+{27 \xi^2 \over \omega^2} \right) \right]
\bar{f}
\end{array}
\eeq
Next, we obtain their critical points
(which are zeros of the beta functions \req{betasgoxf}) by numerical
methods. Note \eg the difference between the $\bar{f}$-coefficients
in $u_{\bar{f}}$ without QG and in the above expression. Is is not
difficult to realize that the QG contributions may get to change
the balance which makes $\Delta$ negative or positive, bringing about
modifications in the limits of the AF region, as we shall see below.

Once we have the numerical values of all these fixed points, we
classify them according to the criterion of the eigenvalues of the
matrix
\beq
\left(
\begin{array}{cccc}
\ds{\partial^2 u_{g^2}     \over \partial(g^2)^2}&
\ds{\partial^2 u_{g^2}     \over \partial g^2\partial\omega}&
\ds{\partial^2 u_{g^2}     \over \partial g^2\partial\xi}&
\ds{\partial^2 u_{g^2}     \over \partial g^2\partial\bar{f}} \\
\ds{\partial^2 u_{\omega}  \over \partial\omega \partial g^2}&
\ds{\partial^2 u_{\omega}  \over \partial\omega^2}&
\ds{\partial^2 u_{\omega}  \over \partial\omega\partial\xi}&
\ds{\partial^2 u_{\omega}  \over \partial\omega\partial\bar{f}} \\
\ds{\partial^2 u_{\xi}     \over \partial\xi\partial g^2}&
\ds{\partial^2 u_{\xi}     \over \partial\xi\partial\omega}&
\ds{\partial^2 u_{\xi}     \over \partial\xi^2}&
\ds{\partial^2 u_{\xi}     \over \partial\xi\partial\bar{f}} \\
\ds{\partial^2 u_{\bar{f}} \over \partial\bar{f}\partial g^2}&
\ds{\partial^2 u_{\bar{f}} \over \partial\bar{f}\partial\omega}&
\ds{\partial^2 u_{\bar{f}} \over \partial\bar{f}\partial\xi}&
\ds{\partial^2 u_{\bar{f}} \over \partial\bar{f}^2}
\end{array}
\right)
\label{notHessian}
\eeq
taken at the critical points in question (notice that this is not a
Hessian matrix, but the standard argument for classifying fixed points
leads us to handle it as such). After this numerical work we find that
in all cases where we had AF without gravity we also have some
solution giving AF with gravity. This is easy to understand by examining
the values of the fixed points: of all the solutions found for
each $(N, n_1, n_2)$-combination, there is at least one
whose value of $\xi$ tends to be around $1/6$
and whose value of $\bar{f}$ is fairly close to $\bar{f}_2$
---the AF fixed point without gravity---
which gives AF in the presence of QG.
In addition, we also find cases where QG makes
possible the existence of AF solutions which are banned without QG
(as was observed in \cite{36}). These new situations correspond to
the values of $N, n_1, n_2$ in the following table
\begin{table}
\begin{center}
\begin{tabular}{|r|r|l|}
\hline\hline
$N$&$n_1$&$n_2$ \\ \hline
 7&0&12 \\
  &1& 7 \\
  &2& 2 \\
 8&0&14 \\
  &1& 8 \\
  &2& 2 \\
 9&0&16 \\
  &1& 9 \\
  &2& 2 \\
10&0&18 \\
  &1&10 \\
  &2& 2 \\ \hline \hline
\end{tabular}
\caption{Values of $N$, $n_1$ and $n_2$.}
\end{center}
\end{table}
and depict a shift in the boundary of the $(N, n_1, n_2)$-region
for which AF existed without QG. More precisely, it is a shift by
decreasing the
allowed value of $n_2$ in one unit, as already commented in \cite{36}.
Thus, we have shown that the $O(N)$ GUT under discussion, interacting
with quantum $R^2$-gravity, may be considered a completely
asymptotically free theory (for some given field contents of this
model). This study has been carried out by introducing
a potential in the space of RG couplings, similar to a $c$-function.

%\subsection{Special cases}
%Since the above discussion on the $(n_1, n_2)$-boundary was restricted
%to $N > 2$, some words about the $N=1,2$ cases are called for (all the
%more reason when this detail was not remarked in \cite{36}).
%
%Stability reasonings based on eigenvalues keep their general validity.
%For $N=1$,
%$\Delta=b^4$ and therefore $\bar{f}_1=0$ and $\bar{f}_2=-b^2/3$. What
%makes this case different from the rest
%is the $b^2 > 0$ condition, which now leads to
%\[ n_1 \geq n_2+3 . \]
%Thus, in the absence of QG, there is AF for all the subcases
%fulfilling this inequality,
%\eg $(n_1, n_2)=(3,0), (4,0), \dots$, $(4,1), (5,1), \dots$, etc.
%When switching on gravity, we find that the same subcases also produce
%sets of fixed points for the enlarged system of couplings.
%However, by looking
%at the eigenvalues of \req{notHessian} at the fixed points in question,
%we realize that they have mixed signs, \ie stability has been lost.
%
%The $N=2$ case is much easier to analyse: given that $b^2=-1/6-4/3n_2$
%this coupling is negative and AF in $g^2$ always fails.

\section{Running gravitational coupling in asymptotically free $O(N)$
GUT with quantum $R^2$ gravity}

Now, having at hand the asymptotically free $O(N)$ GUT interacting
with quantum $R^2$ gravity \req{L25} we may discuss the behaviour
of the gravitational coupling constant. Unlike for GUTs on classical
gravitational backgrounds, we cannot analytically solve the RG
equations for the gravitational coupling constant. Instead, we have
a system of RG equations for the massive couplings
$m^2$, $\gamma\equiv 1/(16\pi G)$ and $\Lambda$, which may be analysed
only numerically (after the corresponding study has been done
for the massless coupling constant). This system may be explicitly
written using the calculations of one-loop counterterms
for massive scalars interacting
with $R^2$ gravity in Ref. \cite{38} and the calculation of the scalar
$\gamma$-function in Ref. \cite{36}.
For the Lagrangian \req{L25} we obtain
%By considering these parts, we are led to study the system
%\req{betaslgoxf} supplemented with the new equations
\beq
\begin{array}{llllrl}
\ds{d\Lambda\over dt}&=&\beta_{\Lambda}&=&\kappa&\ds\left[
{N \over 4}\gamma m^4 +\Lambda N m^2 \left( \xi -{1 \over 6} \right)
+{\lambda\Lambda\over 2}
\left( {43 \over 3}+ {20 \over 3}\omega +{1 \over 6\omega} \right) \right. \\
&&&&&\ds\left. +\lambda^2\gamma\left( {5 \over 4}+{1 \over 16\omega^2}
\right) \right], \\
\ds{d \gamma\over dt}&=&\beta_{\gamma}&=&
-\kappa&\ds\left[ \lambda\gamma\left(
{10 \over 3}\omega -{1 \over 4\omega}- {13 \over 6}
\right)
+N m^2 \left( \xi -{1 \over 6} \right) \right] , \\
\ds{dm^2\over dt}&=&\beta_{m^2}&=&\kappa&\ds\left[
m^2 \left( {N+2 \over 3} f-{3 \over 2}(N-1)g^2 \right) \right. \\
&&&&&\ds +\lambda m^2 \left(
-{43 \over 6}-{5 \over 12\omega}+3{\xi\over\omega}
-{9  \over 4\omega}\xi^2 -{3 \over 2}\xi^2 \right) \\
&&&&&\ds\left. +{\lambda^2 \xi\gamma}
\left( 5+{1 \over 4\omega^2}\xi^2 +{33 \over 2 \omega^2} \right)
\right] .
\end{array}
\label{betasllm}
\eeq

\subsection{Without gravity}
Ignoring all the QG pieces, and performing the changes \req{changes},
we are posed with a system of differential equations consisting of
\req{betasgf} plus
\beq
\begin{array}{lll}
\ds{d\Lambda\over d\tau}&=&\ds
{N \over 4}{m^4 \over g^2}\gamma
+ \Lambda N{m^2 \over g^2} \left( \xi -{1 \over 6} \right), \\
\ds{d \gamma\over d\tau}&=&\ds-N{m^2\over g^2}
\left( \xi -{1 \over 6} \right), \\
\ds{dm^2\over d\tau}&=&\ds m^2
\left[ {N+2 \over 3} \bar{f}-{3 \over 2}(N-1) \right] .
\end{array}
\label{betaslm}
\eeq
%Actually, the simultaneous vanishing of all these beta
%functions leads in general to incompatibility. Here, there are
%difficulties:
%
%\begin{enumerate}
%
%\item
%The zero of the first equation in \req{betasgf} is $g^2=0$, while the
%first in \req{betaslm} gives rise to ${m^4 \over \gamma g^2}=0$.
%
%\item
%From the second equation in \req{betasgf} we have
%got that $\bar{f}$ must be one of the values given by
%\req{barf12}, while the vanishing of the second in \req{betaslm}
%yields either $m^2=0$ or
%\beq \bar{f}_3= {9 \over 2} {N-1 \over N+2} . \label{barf3} \eeq
%Except for $N=1$ ---in which case $\bar{f}_1 = \bar{f}_3=0$---
%$\bar{f}_{1 \atop 2}$ and $\bar{f}_3$ do not exactly
%coincide, and the whole system does not admit simultaneous solution.
%Nevertheless, there can be approximate coincidences. After examining
%$0 \leq n_1, n_2 \leq 20$ and $1 \leq N \leq 1000$, the closest
%similarity we have found takes place at $N=18, n_1=2, n_2=16$, where
%we obtain
%\[ \bar{f}_{1}=3.8278, \ \bar{f}_{3}=3.8250. \]
%(in the other situations studied, the differences are
%$\sim 2\cdot 10^{-2}$ or larger).
%
%\end{enumerate}
%
%In view of this, the only way out seems to be $m^2=0$, and thus we are
%back in the situation in the previous section.

It is possible to estimate the type of $\gamma$ solution to this
system in the large-$\tau$ regime. Considering the most interesting case
\ie $\Delta>0$, ---and therefore $\bar{f}(\tau)$ given by
\req{ftauDeltag0}--- we take approximations of the type
$e^{\sqrt{\Delta}\tau}-1 \sim e^{\sqrt{\Delta}\tau}$ and,
using \req{g2tau}, arrive at
\beq
\begin{array}{llllll}
m^2(\tau)&\sim &\ds m^2(0)e^{\ds -{\cal B}\tau/2},&
{\cal B}&=&\ds {N+2 \over N+8}[ b^2-3(N-1) ]+3(N-1) , \\
\gamma(\tau)&\sim &\ds\gamma(0)
+\gamma_1\left( 1- e^{\ds (b^2-{\cal B}/2) \tau } \right),&
\gamma_1&=&\ds{ N\left( \xi -{1 \over 6} \right) \over
b^2-{\cal B}/2 } {m^2(0) \over g^2(0)}.
%\\
%\Lambda(\tau)&\sim &\ds\Lambda_{\infty}
%-\Lambda_1 e^{\ds -{{\cal B}\over 2\kappa}\tau },&
%\Lambda_1&=&\ds {N m^4(0) \over 2 {\cal B} \gamma(0) g^2(0)} .
\end{array}
\eeq
Actually, ${\cal B}$ is
positive whenever we are in the settings of AF for both
$g^2$ and $\bar{f}$ ---\ie $(N, n_1, n_2)$ combinations in the
first table--- while $b^2-{\cal B}/2$ is negative in these
same cases.
As we see, under the present
assumptions $\gamma(\tau)$ would tend asymptotically to a constant
value of $\gamma(0)+\gamma_1$ as $\tau\to\infty$.
Undoing now the variable change \req{changes}, we may put
$\tau={1 \over b^2}\log( 1+\kappa b^2 g^2(0) t )$, and writing
$G(t)$ in terms of $\gamma(t)$ we are left with
\beq
G(t)=G(0)\left\{ 1-16\pi G(0) \gamma_1
\left[ ( 1+\kappa b^2 g^2(0) t )^{1-{{\cal B}\over 2b^2}} -1 \right]
\right\}^{-1} ,
\eeq
where, obviously, $G(0)=1/(16\pi\gamma(0))$. This expression coincides
with \req{538} after making the notational replacements
$N_s \to N$, $B^2\to b^2$, $2b+1 \to 1-{{\cal B}\over 2b^2}$.
The above remarked asymptotic behaviour is clearly appreciated
%for $\gamma(t)$
in Fig. 1.

\subsection{With Gravity}

%Numerical integration of the complete system of differential equations
%shows $\Lambda(t)$ increasing without bound as $t\to\infty$.
%Therefore, should we throw this model into the basket?
We can solve numerically the whole system of differential equations and
examine the asymptotically free regime of its solutions. By plotting
the gravitational running coupling, we obtain the curves $a$ and $b$ in
Fig. 2.

\section{Gravitational field equations with GUT quantum corrections in
curved spacetime}

Let's now turn to some other application of the running coupling
constants in curved spacetime, namely to the RG-improved effective
Lagrangian \req{Leff533}. We will work for simplicity on the purely
gravitational (almost constant) background, supposing that for all
quantum fields we have zero background. In actual one-loop calculations,
it turns out that, working in configuration space, the RG parameter $t$
is typically of the form
\beq
t\simeq {1 \over 2}\log{c_1 R+ c_2m^2 \over \mu^2 } ,
\label{formoft}
\eeq
where $c_1$, $c_2$ are some numerical constants, and $m^2$ is the
effective mass of the theory. In the model under discussion we have a
few different masses, so there is no unique way of choosing only one
functional form for $t$ \cite{27,FoJoStEi}. Hence, we will consider the
regime of strong curvature when curvature is dominant in
\req{formoft}. Then, the natural choice of $t$ in the RG-improved
Lagrangian is (see also \cite{27,26})
\beq
t ={1 \over 2}\log{R \over \mu^2 }.
\eeq
Note that such a regime may lead to curvature-induced asymptotic freedom
\cite{26}. In this regime we obtain that expression (\ref{Leff533})
gives the leading-log approach to the whole perturbation series.
\beq
\begin{array}{ll}
S_{\mbox{RG}}=\int d^4 x \sqrt{-g}&\ds\left\{
\Lambda(t)-{1 \over 16\pi G(t)}R \right. \\
&\ds\left. +a_1(t)R^2+a_2(t)C_{\mu\nu\alpha\beta}C^{\mu\nu\alpha\beta}
+a_3(t)G \right\} .
\end{array}
\label{SRG}
\eeq
We will consider only asymptotically free GUTs, where $\Lambda(t)$,
$G(t)$ are given in Sect. 2,
$t ={1 \over 2}\log{R \over \mu^2 }$,
$a_{2,3}(t)=a_{2,3}+\tilde a_{2,3}t $ (see \req{betasaslg}). Note,
however, that equation \req{SRG} is very general. In particular,
it has a similar form in the asymptotically free GUTs with $R^2$-gravity
of Sects. 6,7 in strong curvature regime (however, the $t$-dependence of
the running couplings $\Lambda(t), \dots, a_3(t)$ is of course
different). Notice that the technique developed in Ref. \cite{CaLoAu}
may be very useful for the explicit solution of quantum corrected
gravitational field equations for non-constant curvature.

Working in constant curvature space
\beq
R_{\mu\nu}={ g_{\mu\nu}\over \beta^2}, \ \  R={4 \over \beta^2}
\eeq
we may rewrite \req{SRG} expanding the coupling constants
up to linear $t$ terms as follows
\beq
\begin{array}{rl}
S_{\mbox{RG}}= \mbox{const}\times \beta^4 &\left\{
\Lambda +\tilde\Lambda t \ds
-{1 \over 16\pi}\left( {1 \over G} + \tilde G t \right) {4 \over \beta^2}
\right. \\
&\ds\left. +(a_1 + \tilde a_1 t ){16 \over \beta^4}
+(a_3+\tilde a_3 t) {8 \over 3\beta^4} \right\}, \\
t&\ds={1 \over 2}\log{4 \over \beta^2 \mu^2 },
\end{array}
\eeq
where the explicit form of the constants $\tilde\Lambda$, $\tilde G$, $\tilde
a_1$,
$\tilde a_3$ is evident from \req{betasaslg}.

Now one can write the field equation which fixes $\beta^2$ in terms of the
theory
parameters:
\beq
\begin{array}{ll}
\ds{\partial S_{\mbox{RG}} \over \partial\beta^2}=
\mbox{const}\times&\ds \left\{
\left[
-{\tilde\Lambda\over 2}+2(   \Lambda +\tilde\Lambda t ) \right] \beta^2
 -{1 \over 4\pi}  \left[ -{\tilde G \over 2}
+ \left( {1 \over G} + \tilde G t \right)  \right]  \right.  \\
&\ds\left.
-\left( 16 \tilde a_1 + {8 \over 3}\tilde a_3 \right){1 \over 2 \beta^2}
                \right\} =0.
\end{array}
\label{dSRG}
\eeq
One can see that there exists a real root of \req{dSRG} for most asymptotically
free GUTs (for a massless theory, see also discussion in Ref. \cite{26}).
Hence, we have
got a non-singular universe with a metric of the form
\beq
ds^2 = a^2(\eta)\left(
d\eta^2 -{ dx^2+dy^2+dz^2 \over
\left( 1 +  { k r^2 \over 12} \beta^2  \right)^2 } ,
\right)
\label{metric}
\eeq
where $k=1,0,-1$ for a closed, asymptotically open and open universe,
respectively,
and
\beq
a(\eta)={\sqrt{k}\over 2}\left\{
{\rm tg}\left( -{\sqrt{k} \eta \over 2 \sqrt{3} \beta } \right)
+{\rm tg}^{-1}\left( -{\sqrt{k} \eta \over 2 \sqrt{3} \beta } \right)
\right\}
\eeq
as the solution of the gravitational field equations with quantum GUT
corrections.
The GUTs quantum corrections are here present in the form of $\beta$, which is
defined
by GUT parameters from \req{dSRG}. In terms of physical time $T$, the scale
factor may be put as
\beq
a(T)= {\sqrt{3} \beta \over 2} \left[
e^{T /(\sqrt{3} \, \beta)}
+k e^{ -{T /( \sqrt{3} \, \beta)} }
\right] .
\label{aT}
\eeq
Note that such a non-singular inflationary type (De Sitter) solution in gravity
theories with higher derivative terms (induced by quantum matter) has
been discussed
in Refs. \cite{31,26,Star,CaLoAu} in different contexts (for a very recent
discussion
on non-singular cosmologies in higher-derivative theories,
                                                  see \cite{BrMuSo}).

Let us consider the closed universe ($k=1$) in  \req{metric}. In this
case, for
some choices of the theory parameters one can find an imaginary $\beta$ as the
solution of Eq. \req{dSRG}. Similarly to what happens in papers \cite{GiSt},
such a solution may be interpreted as a Lorentzian wormhole which connects two
De Sitter universes. However, this wormhole, that
for $| \beta^2 | \sim L_{Pl}^2$ has a `mouth' of the Planck size,
results from quantum GUT corrections.  From another viewpoint, working
with \req{aT} at $k=1$ and imposing the initial conditions $a(0)=R_0$,
$\dot{a}(0)=0$ we get
\beq
a(T)= \sqrt{3}\beta \cosh\left( T \over \sqrt{3} \beta \right) .
\label{aTcosh}
\eeq
Some analysis of this solution is given in \cite{Be}. For example, when the
effective cosmological constant is zero, the solution \req{aTcosh}
corresponds to a closed universe connected through a wormhole to flat space.
In a similar fashion one can construct the quantum corrected gravitational
equations in other regimes and study their solutions.

\section{Discussion}

In the present paper we have discussed the running of the gravitational
coupling
constant in asymptotically free GUTs in curved spacetime, in the effective
theory
for the conformal factor and in asymptotically free $R^2$-gravity interacting
with an $O(N)$ GUT. The running gravitational coupling constant has been used
to calculate the leading quantum corrections to the Newtonian potential.
These corrections have logarithmic form in asymptotically free GUTs and in
the effective theory for the conformal factor, and power-like form
(but of different nature from that in Einsteinian gravity) in finite
GUTs.
In $R^2$-gravity with  $O(N)$ GUT, the behaviour of the gravitational coupling
constant is numerically analysed. Its decay rate gets higher as the value of
$m^2(0)$
is raised.

The running coupling constants are also necessary in other respects,
particularly
in the RG-improvement procedure. We have found the RG-improved effective
gravitational
Lagrangian in the regime of strong constant curvature, and have discussed the
non-singular De Sitter solution of the corresponding quantum corrected
gravitational
field equations. The present technique is quite general and can be applied in
various
situations, in particular for the construction of RG-improved non-local
gravitational
Lagrangian, what we plan to discuss elsewhere.

The other interesting field where the running gravitational constants
calculated
in this paper play an important role is in the quantum corrections to
the Hawking-Bekenstein black hole entropy.
The black hole entropy (Bekenstein-Hawking formula) has the
following form
\[
S=\frac{A}{4G},
\]
where $A$ is the surface area of the event horizon and $G$ the

gravitational constant. It has been suggested by Susskind et al.
\cite{30} that
in the brick wall approach (e.g. the 't Hooft cut-off  regularization
\cite{47n}) the quantum corrections to the black-hole entropy can be
absorbed in the above formula as a simple renormalization of the
gravitational constant $G$. In this respect, taking into account the
results of our study we are led to conjecture that in the case of
dimensional regularization (where only logaritmic divergences are
important) the quantum corrections to the Beckenstein-Hawking formula
are precisely given by the standard renormalization of $G$, as discussed
above in the present paper ---plus some less essential contributions
from the higher-derivative terms in (2.1). This gives further
relevance to our calculations in the paper, since then, in order to
take into
account GUT contributions to the black hole entropy one just has to use
$G(t)$ instead of $G$ in the formula for the entropy.

Note that
the quadratic terms in the lagrangian contribute the black hole
entropy already at tree level as shown
in Refs \cite{EcLo,CaLoAu,CaLoAu95}.
For the
charged black hole they give corrections like
$$
S={A_H\over 4G}+{16\pi^2(3a_2+2a_3) Q^2\over A_H}+...
$$
 From the results of our paper we can
futher infer that the one-loop
corrections will have the form of this equation but with all
the coupling constants being now a function of t. We hope to
return to this question in near future.

\vspace{5mm}

%\newpage

\noindent{\large \bf Acknowledgments}

We would like to thank I.L. Shapiro for discussions on Sect. 7.
SDO thanks N.P. Chang for some remarks.
AR is grateful to J. Isern for an enlightening discussion.
This work has been supported by DGICYT (Spain), project Nos.
PB93-0035 and SAB93-0024, by CIRIT (Generalitat de Catalunya)
AEN 93-0474, and by RFFR, project 94-020324.

\newpage
%\vskip1cm
\ni{\large\bf Figure captions}

\ni{\bf Fig. 1.} Running gravitational coupling
$G(t)=1/(16\pi\gamma(t))$ obtained by
numerical integration of the full system made of Eqs.
\req{betasgoxf} (written in terms of the original RG parameter $t$
of Ref. \cite{36}) and Eqs. \req{betasllm}, for $N=7$, $n_1=0$,
$n_2=13$. The initial values
are $g^2(0)=0.1$, $\omega(0)=0.1$, $\xi(0)=0$, $f(0)=0.5$,
$\Lambda(0)=0.1$ ,$\gamma(0)=0.1$, $m^2(0)=0.1$.
and $\lambda(0)=0$, \ie initially there is no quantum gravity.
The asymptotic tendency of $G(t)$ towards a constant value
is clear.

\ni{\bf Fig. 2.} $G(t)$ in presence of QG
with the same initial conditions
as in Fig. 1, except for: $a$) $\lambda(0)=0.5$ and $m^2(0)=0$,
$b$) $\lambda(0)=0.5$ and $m^2(0)=0.5$. The running gravitational
constant is quickly decreasing.

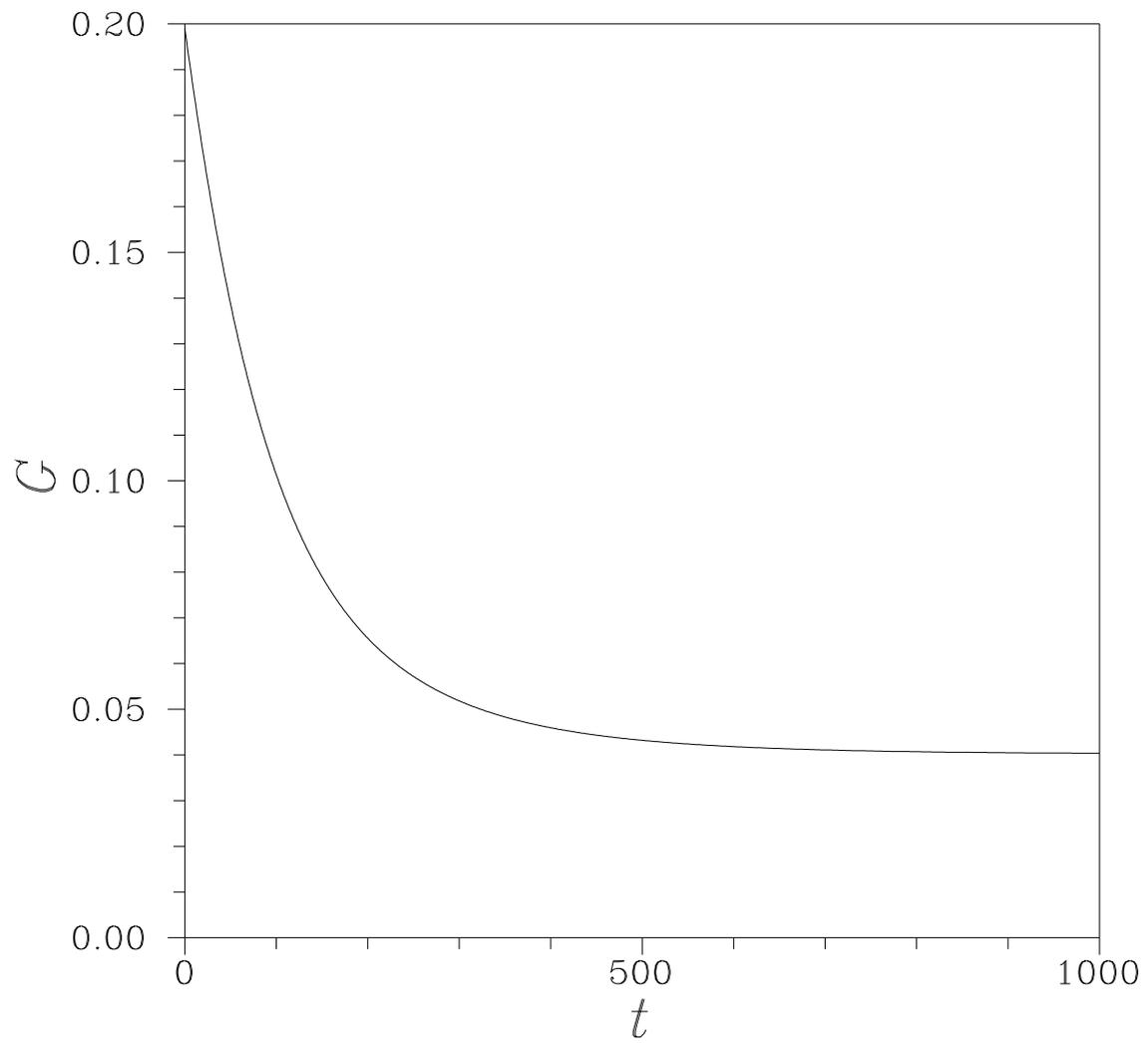
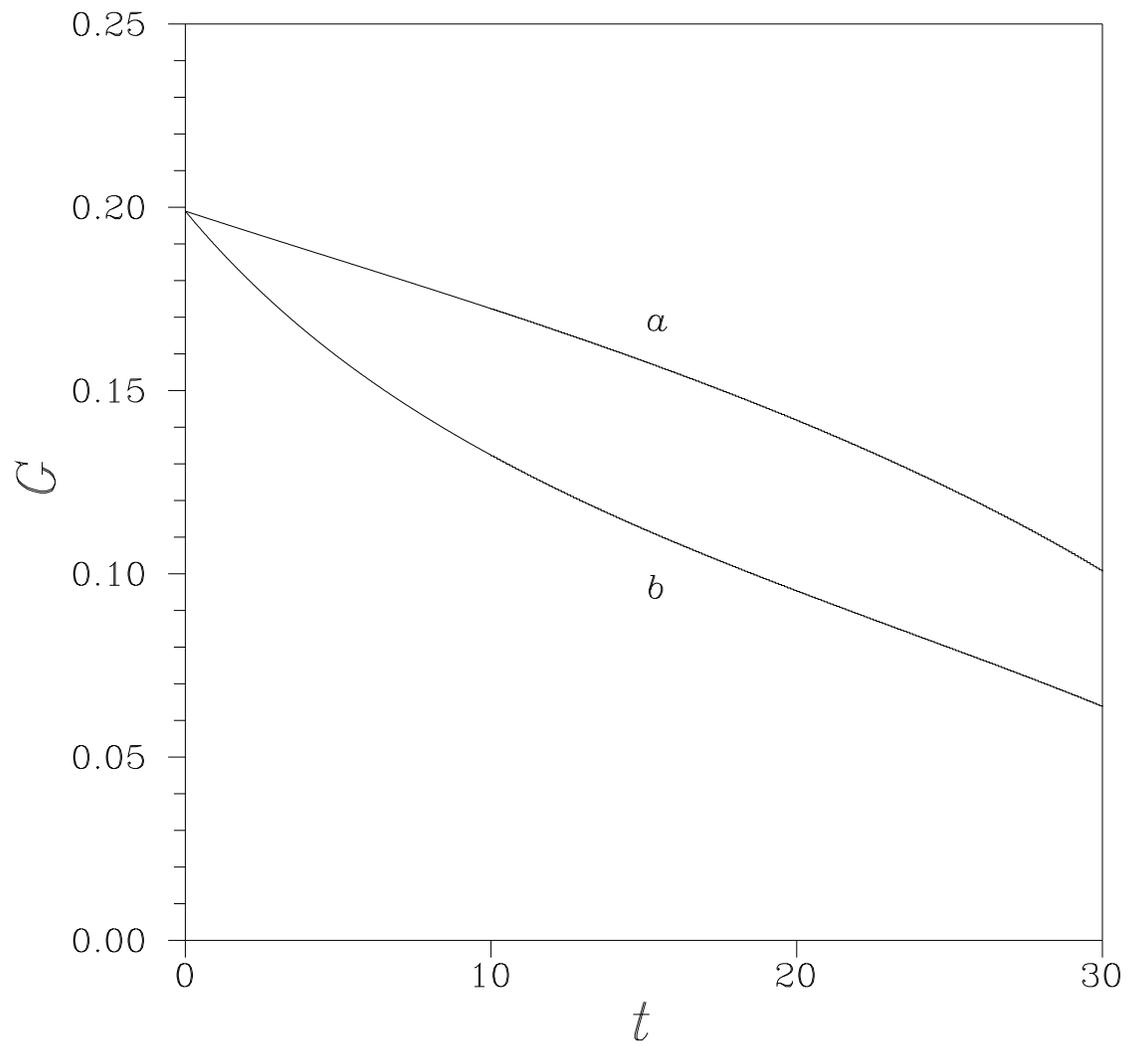
\end{document}